\documentclass[prl, twocolumn, amsmath, amSymb]{revtex4}
\usepackage{dcolumn}% Align table columns on decimal point
\usepackage{bm}% bold math
\usepackage{graphicx}% Include figure files
\usepackage{color}
\usepackage{subfigure}
\usepackage{hyperref}
\usepackage{latexsym}
\usepackage{amsthm}
\usepackage{amssymb}
\DeclareGraphicsExtensions{.jpg,.pdf, .mps, .png, .eps, .ps, .EPS,.gif}
\begin{document}

\def\bea{\begin{eqnarray}}
\def\eea{\end{eqnarray}}

\newcommand{\avg}[1]{\langle{#1}\rangle}
\newcommand{\Avg}[1]{\left\langle{#1}\right\rangle}

\title{Enhancing the robustness of a multiplex network leads to \\ multiple discontinuous percolation transitions}

\author{Ivan Kryven}\email{i.kryven@uu.nl}
\affiliation{Mathematical Institute, Utrecht University, PO Box 80010, 3508 TA Utrecht, the Netherlands}
\author{Ginestra Bianconi}
\affiliation{
School of Mathematical Sciences, Queen Mary University of London, London, E1 4NS, United Kingdom \\
The Alan Turing Institute, the British Library, London NW1 2DB, United Kingdom
}
\begin{abstract}
Determining design principles that boost robustness of interdependent networks is a fundamental question of engineering, economics, and biology. It is known that maximizing the degree correlation between replicas of the same node leads to optimal robustness. Here we show that increased robustness might also come at the expense of introducing multiple phase transitions. These results reveal yet another possible source of fragility of multiplex networks that has to be taken into the account during network optimisation and design. 
\end{abstract}

\maketitle

Multilayer networks \cite{bianconi2018multilayer,PhysReport,Kivela} formed by several interacting layers have emerged as a powerful framework to analyse a  variety of complex systems, including such classical examples as global infrastructures, economic networks, temporal networks and the multi-level structure of the brain. Other disciplines, as material science and chemical synthesis, are on the way to adopt the network science toolbox, less for analysis purpose, but mainly for its potential for optimisation and rational design \cite{schamboeck2019,kryven2018a,orlova2018automated,kryven2016,Kryven2016b}.
Therefore, predicting the robustness of multilayer networks \cite{Havlin,Baxter,Son}, assessing the risk of large avalanches of failures \cite{Havlin,Baxter,Son,Havlin2,Kahng1,Kahng2,rapisardi2019fragility}, and designing robust multilayer architectures \cite{Goh,Makse,Redundant} are the key theoretical questions entailing implications for engineering, economics, and biology. 
Recent works on percolation in multilayer networks revealed that the correlation between intra-layer degrees of replica nodes \cite{Goh,Makse} and link overlap \cite{cellai2013,Goh_viability,cellai2016b} have profound consequences in determining the response of a multiplex network to random damage. The case of positive inter-layer degree correlation \cite{Goh,bianconi2018multilayer}, when a hub node in one layer is likely to be interdependent on a hub node in another layer, is known to increase robustness of interdependent multiplex networks. It is widely believed that maximizing the inter-layer degree correlation is a good design principle for building robust infrastructures. The network optimization is used not only for engineered networks and infrastructures \cite{aldous2019optimal,valente2004,paul2004,paul2005,tanizawa2005,tanizawa2006} but also for economic \cite{guimera2002optimal} and biological networks \cite{sole2003optimization,thai2011handbook}.
 Therefore, it is of fundamental importance to understand how maximizing the inter-layer degree correlation may affect the properties of multiplex networks.

In this paper we demonstrate that such an optimization strategy might come at a cost: unexpectedly,
multiplex networks with strong intra-layer degree correlations may be prone to multiple percolation transitions.  Several works already showed that percolation processes on interdependent networks may be associated with multiple phase transitions \cite{bianconi2014,colomer2014,kryven2017,costa,kryven2019}.
For instance, when an interdependent multilayer network is drawn from the configuration model, percolation may lead to multiple discontinuous and hybrid transitions due to a successive deactivation of different layers at different values of the percolation parameter \cite{bianconi2014}.
Similar effect is also seen in classical percolation as it can display multiple phase transitions corresponding to deactivation of one or several layers at a time \cite{colomer2014,kryven2019}. Nevertheless, these phase transitions are continuous and second order. Here we show that interdependent multiplex networks may be decomposed during percolation by multiple discontinuous phase transitions into sub-multiplexes that, to the contrary to previous observations, span across all layers.
%In these cases percolation progressively strips a multiplex network from the nodes with degree less than a given threshold.  

%%% FIGURE 1 %%%
\begin{figure}
\begin{center}
\includegraphics[width=\columnwidth]{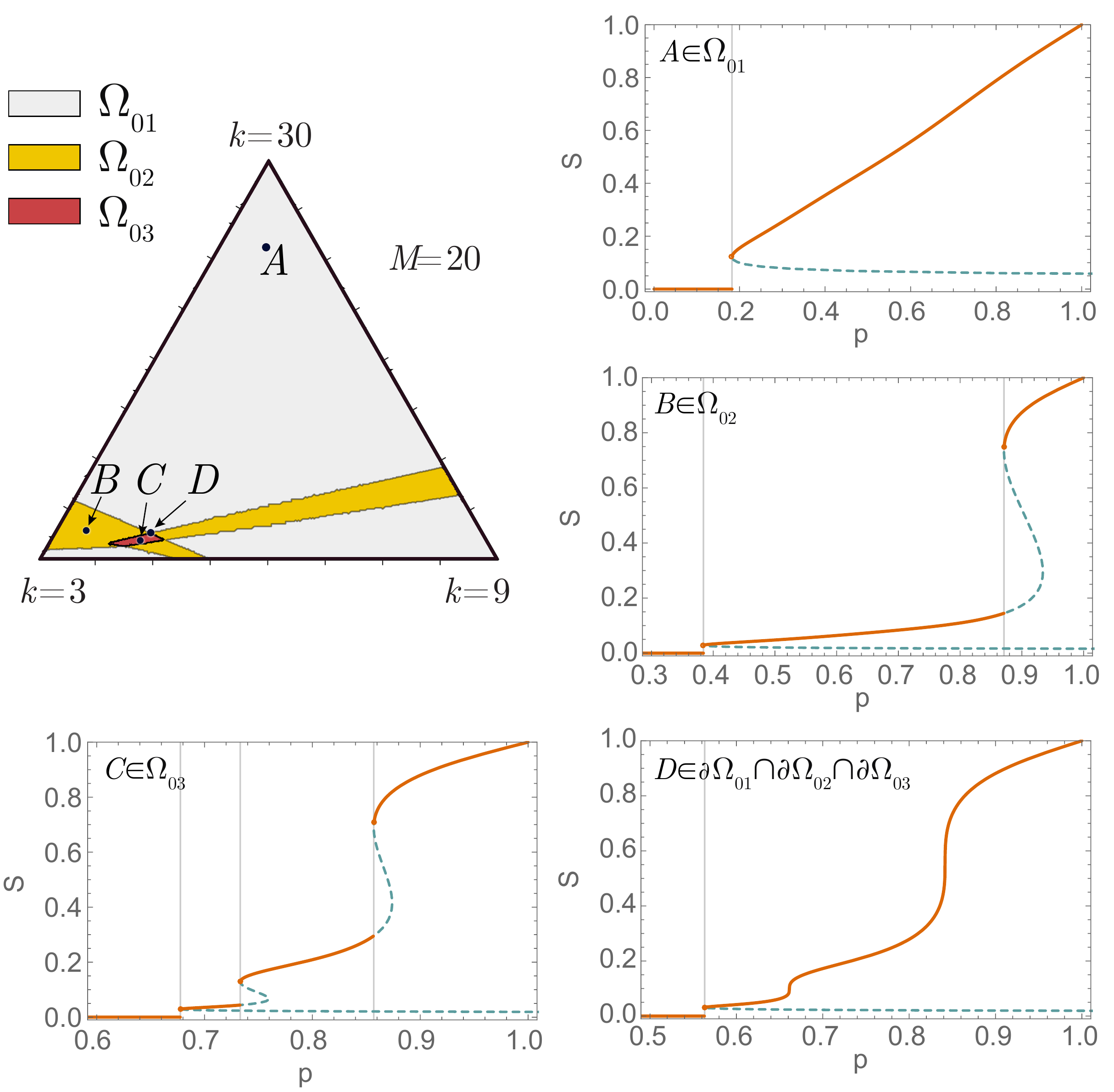}
\caption{ The phase diagram for interdependent percolation in multiplex networks with degree-activity distribution $P(k,B)=\delta_{B,20} \hat{P}(k),$ and $\hat{P}(k)=c_1\delta_{k,3} + c_2\delta_{k,9}+c_3\delta_{k,30}$ is depicted with a barycentric plot featuring domains $\Omega_{ij}$ with $i$ continuous and $j$ discontinuous phase transitions.  \emph{Panels:} The fraction  $S$ of  nodes in the Mutually Connected Giant Component (MCGC) for the points $A,B,C,D$, of  barycentric coordinates ($c_1,c_2,c_3$):  $A=(0.095,0.095,0.810)$; $B=(0.87,0.04,0.09)$; $ C=(0.75,0.2,0.05)$, and  $D=(0.73,0.20,0.065)$. The dashed lines indicate the unstable branch and the vertical guidelines indicate the predicted positions of the discontinuous phase transitions.
 \vspace{-0.8cm}
 }
\label{fig:1}
\end{center}
\end{figure}
%%% FIGURE 2 %%%
\begin{figure*}[htbp]
\begin{center}
\includegraphics[width=\textwidth]{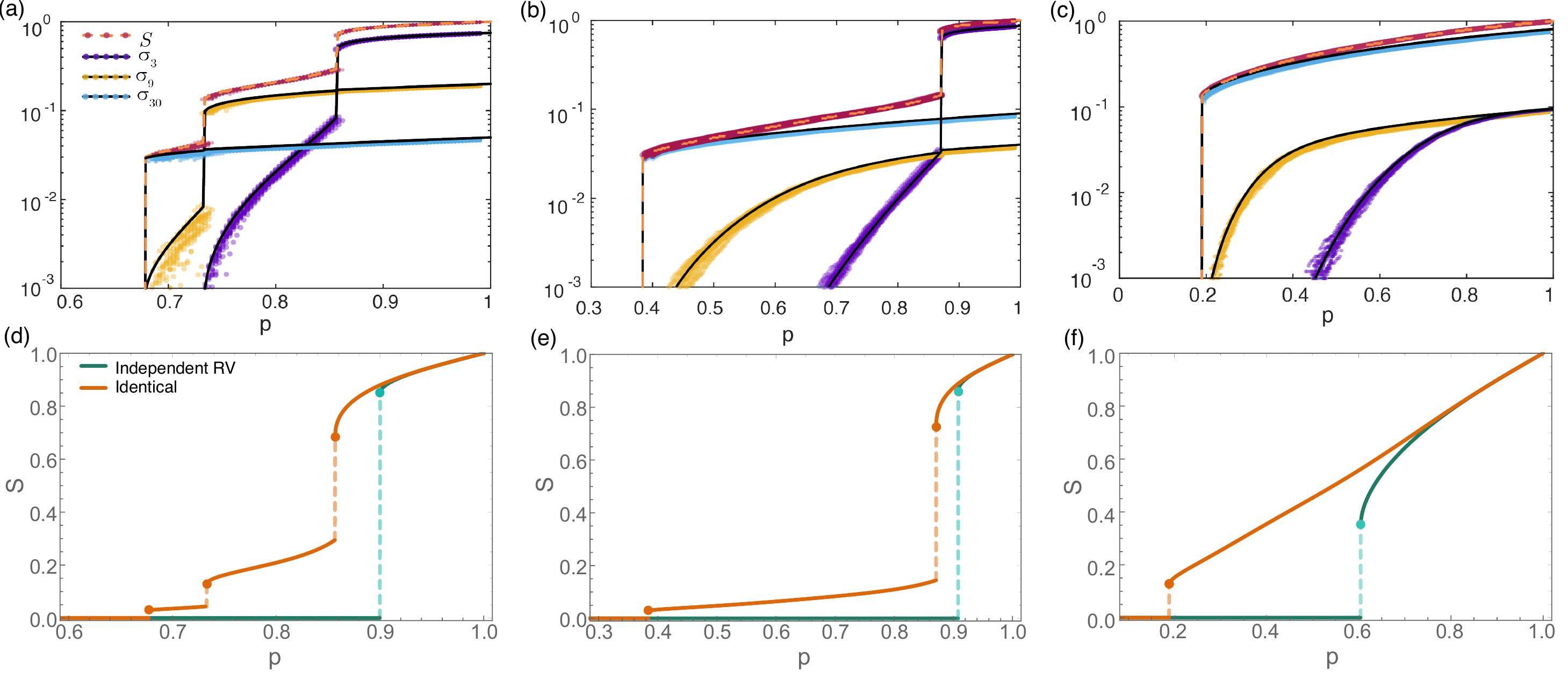}
\caption{ Evolution of the size and structure of MCGC during interdependent percolation.
 (a,b,c) Theoretical predictions (\emph{solid lines}) and simulations results (\emph{dots}) for size $S$ of MCGC and the fraction $\sigma_{k}$  of nodes that have intra-layer degree $k$ are plotted at different points of the phase diagram from Fig.~$\ref{fig:1}$: $C$ -- panel (a), $B$  -- panel (b) and $A$ -- panel (c).
(d,e,f) The fractions  $S$ of nodes in the MCGC  (\emph{solid lines}) are compared against randomised network in which  replica degrees are independent random variables (RV) (\emph{dashed lines}) and showed at different points from the phase diagram:  $C$ -- panel (d), $B$ -- panel (e), and $A$ -- panel (f).
The simulations in panels (a,b,c) are averaged over 20 realisations of networks with $10^5$ nodes and $M=20$ layers.
  \vspace{-0.5cm} }
\label{fig:simulations_and_independent_degrees}
\end{center}
\end{figure*}

Let $P(k,B)$ be the joint probability that a randomly chosen node has degree $k$ in every layer and activity $B$, \emph{i.e.} it is dependent on $B-1$ replicas in other layers.
We identify three classes of such multiplex networks: 
1. multiplex networks with multimodal degree distribution and a node's replicas having identical degrees in all layers, therefore featuring pairwise degree correlation one,
2. regular multiplex networks with all nodes having identical degrees and multimodal activity distribution.
3. multiplex networks with a multimodal joint degree-activity distribution and identical degrees of a node's replicas.

In this paper multiplex networks are modelled by the maximum-entropy ensemble preserving the degree-activity distribution $P(k,B)$, and therefore, as explained in Appendix A, size $S$ of the  MCGC is found by solving the following equations:
\bea\label{eq:MS}
S&=&p\sum_{k,B>0}P(k,B)[1-(1-s)^k]^{B},
\eea
where $p$ is the probability that the edge is present and $s$ is the fixed point of:
$$
s=p\sum_{k,B>0}\frac{k}{\avg{k}}P(k,B)[1-(1-s)^{k-1}][1-(1-s)^k]^{B-1}.
$$
Interdependent percolation can also be interpreted as the steady state of a interdependent epidemic spreading process \cite{Alex,wang2019,Son,bianconi2018multilayer} in which each node must have at least one infected neighbour in each layer to become infected. Under this interpretation, the discontinuous phase transitions can be interpreted as a successive abrupt invasion of different sub-multiplexes by an epidemic.

\emph{Constant activity and multimodal degree distribution.}
 Consider a multiplex network in which  all nodes have the same activity $B=M$ and 
 the intra-layer degrees are distributed according to the degree distribution $\hat{P}(k)$. The degree-activity distribution is then given by:
\bea \label{eq:PkB1}
P(k,B)=\delta_{B,M} \hat{P}(k),
\eea
where $\delta_{a,b}$ is the Kronecker's delta.
 Let the intra-layer degree distribution be multi-modal, as defined by:
\bea\label{eq:ex1}
\hat{P}(k) = c_1\delta_{k,k_1} + c_2\delta_{k,k_2}+c_3\delta_{k,k_3},
\eea
where the normalisation condition imposes the constraint $c_1+c_2+c_3=1$.
We refer to all combinations of $c_1,c_2,c_3\in[0,1]$ that satisfy above-mentioned normalisation condition as the phase space of the model \eqref{eq:PkB1}. Note that even though we have three parameters to chose, there are only two degrees of freedom.
Networks satisfying equation \eqref{eq:PkB1} arise as a result of top-down design  \cite{valente2004,paul2004,paul2005,tanizawa2005,tanizawa2006}
and appear  in temporal network frameworks \cite{masuda2016guidance}.

In what follows, we report existence of peculiar domains in the phase space, $$\Omega_{i,j}\subset\{(c_1,c_2,c_3): c_1,c_2,c_3\in[0,1]\},$$
at which, generally speaking, the percolation process features $i$ continuous and $j$ discontinuous phase transitions.
When the multiplex network is defined by \eqref{eq:ex1}, and as long as the degrees $k_1,k_2$ and $k_3$ are sufficiently distant and number of layers $M$ is sufficiently large, one observes up to three discontinuous phase transitions, that is $i=0$ and $j=1,2,3$.  The phase diagram of such model can be represented by using the barycentric coordinate system in which each point has coordinates $(c_1,c_2,c_3)$.
By studying the critical behaviour of  Eq. \eqref{eq:MS} for this choice of the degree distribution, we identify three distinct domains in the phase diagram as indicated by $\Omega_{01},\Omega_{02}$ and $\Omega_{03}$, see Fig.~\ref{fig:1}.
Each of the discontinuous phase transitions corresponds to progressive deactivation of distinct sub-multiplexes. Both theoretical predictions and simulations of the percolation process, see Fig.~\ref{fig:simulations_and_independent_degrees}a,b,c, show that these phase transitions correspond to the deactivation of the nodes in the order of their increasing intra-layer degrees. The nodes that are damaged first are the nodes with the lowest intra-layer degree, then the nodes with the second lowest value of the intra-layer degree are damaged, and finally, also the nodes with the largest intra-layer degree fail. Conversely, in the epidemic spreading interpretation of the interdependent percolation, these transitions correspond to an abrupt successive invasion of the epidemics into different sub-multiplexes. Therefore, these three transitions can be revealed by monitoring  the order parameters $\sigma_k$ indicating the fraction of nodes that have  intra-layer degree $k$  and that belong to the MCGC as a function of $p$ (see Fig.~\ref{fig:simulations_and_independent_degrees}a,b,c). 

\begin{figure}
\begin{center}
\includegraphics[width=\columnwidth]{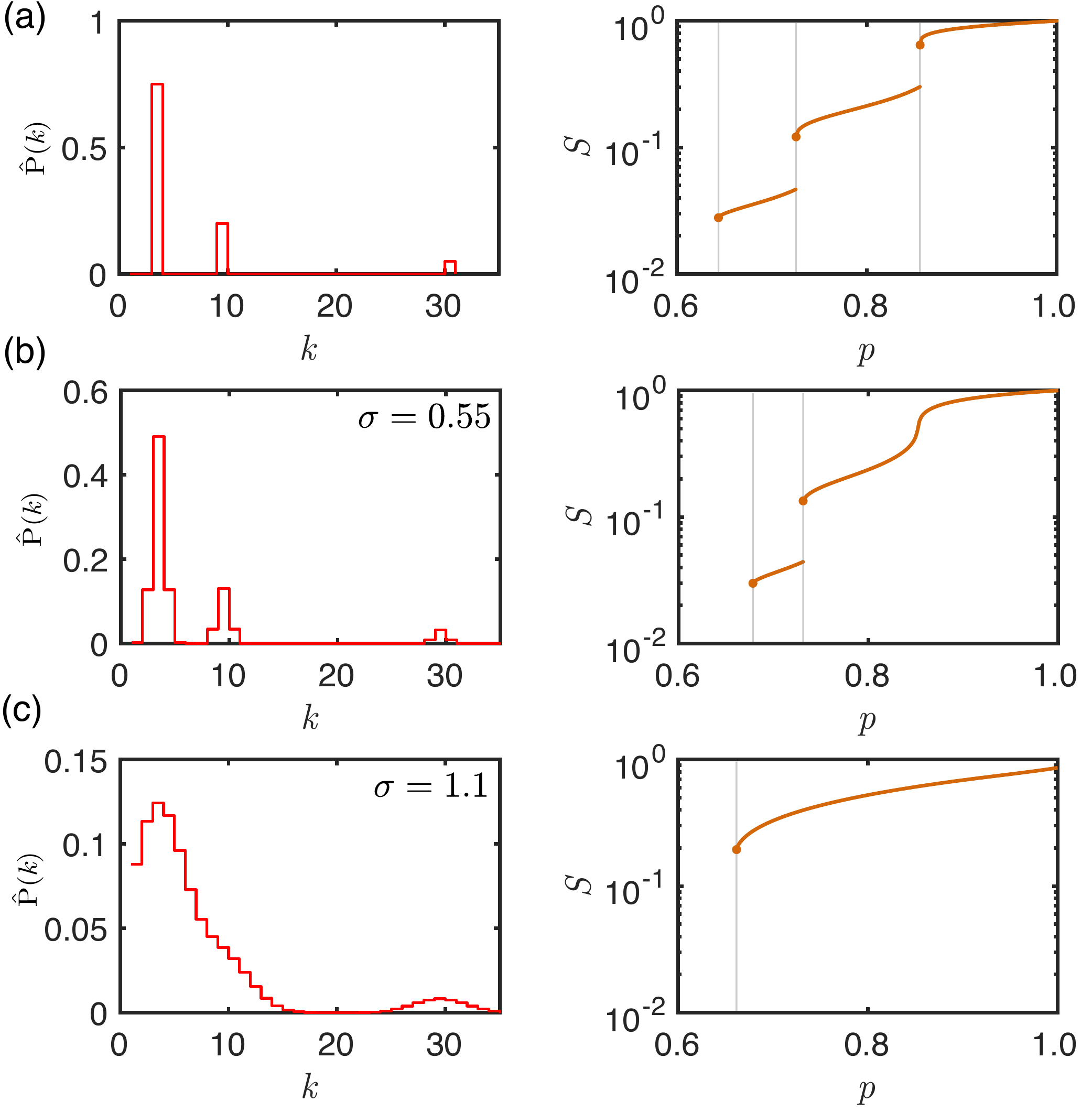}
\caption{ The influence of multimodality of degree distributions on the number of phase transitions.
Three examples of degree distributions given by $P(k,B)=\delta_{B,20}\hat{P}'(k)$  are shown together with the analytically obtained fraction $S$ of nodes in the MCGC. Distributions $\hat{P}'(k)$  are obtained by applying Gaussian filters to 
$\hat{P}(k)=c_1\delta_{k,3}+c_2\delta_{k,9}+c_3\delta_{k,30}$ with $(c_1,c_2,c_3)=(0.75,0.2,0.05)$.	 Case (a) depicts the unmodified $\hat{P}(k)$ distribution; cases (b) and (c) are obtained by applying the Gaussian window filter with standard deviation $\sigma=0.55$  and  $\sigma=1.1$ respectively.
}
 \label{fig.GaussianWindow}
\end{center}
\end{figure}

In order to investigate the effect of the inter-layer degree correlations, we have considered a multiplex network in the region  $\Omega_{03}$ that features three discontinuous phase transitions and compared the robustness of this network with the robustness of a null model, which is formed by a multiplex network having independent degrees of replica nodes, as explained in Appendix B. As can be seen in Fig.~\ref{fig:simulations_and_independent_degrees}d,e,f,  the multiplex networks with identical replica degrees are more robust than the null model, {\it i.e.} for every value of $p$ they have a larger MCGC, which was also reported in Ref. \cite{Goh}. However such an optimization induces two additional discontinuous phase transitions that are not observable in the null model. Therefore, we conclude that optimization of network robustness by maximising the inter-layer degree correlations might come at the expense of inducing additional discontinuous phase transitions. 
The decomposition of the studied multiplex networks into sub-multiplexes indicates that having multiple modes in the degree distribution is necessary to maintain multiple phase transitions. 
The presence of multiple phase transitions is strongly favoured by the multimodality of the degree distribution.
 Figure~\ref{fig.GaussianWindow} applies a series of the Gaussian window filters (\emph{i.e.} discrete convolution with a Gaussian function) with progressively increasing standard deviation to reduce the segregation between the peaks in the case from $\Omega_{03}$ and shows that, as the peaks merge, the corresponding discontinuous phase transitions disappear.  
In overall, the investigations presented in this paper suggest that a large number of layers, segregated multi-modal degree distribution, and identical degrees of nodes' replicas jointly provide sufficient conditions for the presence of multiple discontinuous phase transitions in percolation of interdependent multiplex networks.\\

\begin{figure}[t!]
\begin{center}
\includegraphics[width=1\columnwidth]{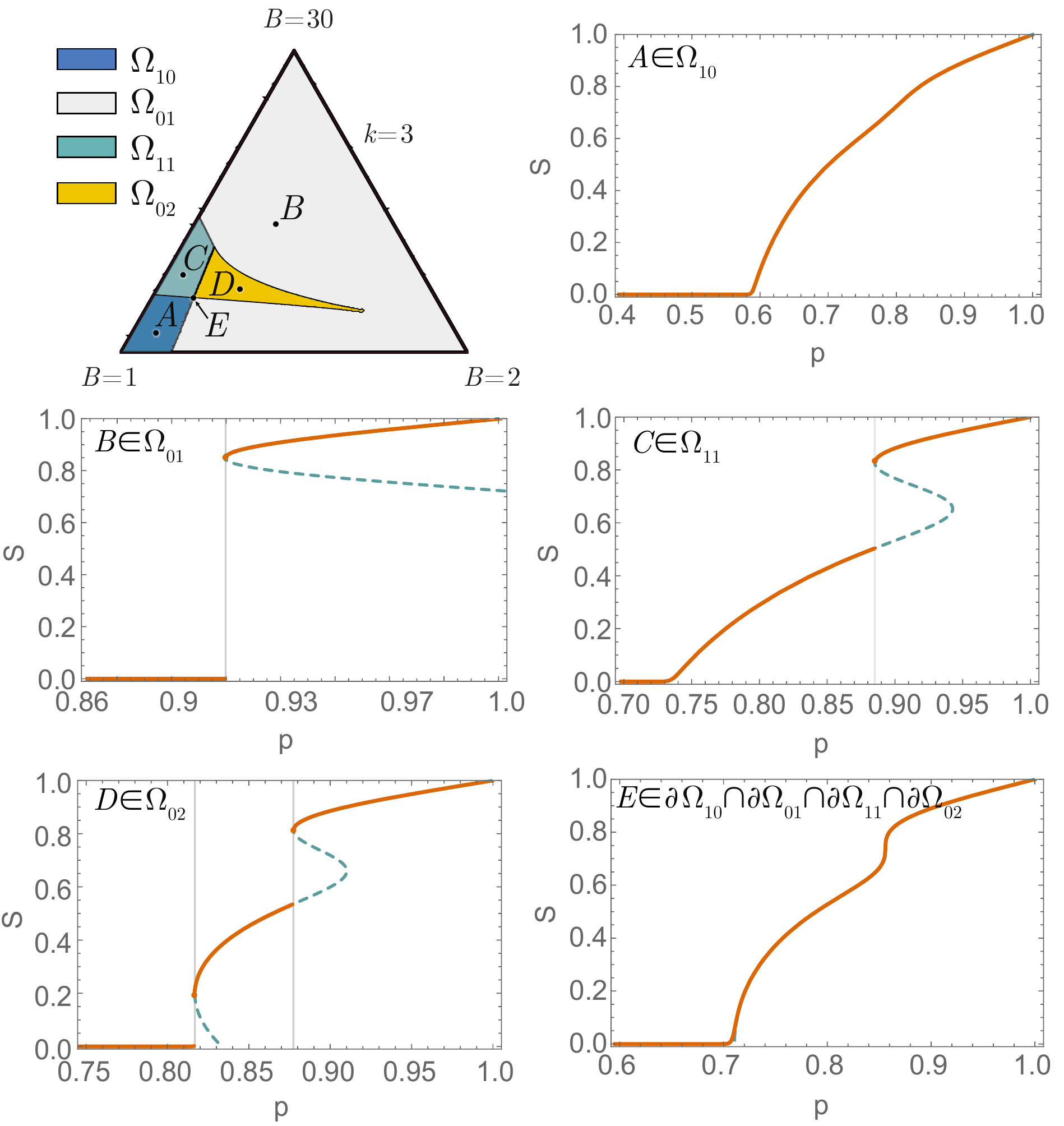}
\caption {
The phase diagram for interdependent percolation in a regular multiplex network
 defined by degree distribution 
$P(k,B)=\delta_{k,3} \tilde{P}(B),$ $\tilde{P}(k)=c_1\delta_{B,1} + c_2\delta_{B,2}+c_3\delta_{B,30}$.
The barycentric plot features domains  $\Omega_{ij}$ with $i$ continuous and $j$ discontinuous phase transitions.   {\it Panels:} The fraction $S$ of nodes in the MCGC for points $A,B,C,D$ and $E$ having the barycentric coordinates $(c_1,c_2,c_3)$: $A=(0.85,0.06,0.09)$, $B=(0.33,0.33,0.33)$, $C=(0.68,0.04,0.28)$, $D=(0.6,0.16,0.24)$, and  the shared accumulation point $E=(0.70,0.11,0.18)$.
\vspace{-0.7cm}
}
\label{fig.SMbarycentricA}
\end{center}
\end{figure}

\emph{Variable activity and constant intra-layer degree.}
 In the previous paragraphs we showed that a multi-modal degree distribution can lead to multiple percolation transitions in multiplex networks with identical intra-layer degrees of nodes' replicas and the identical  activity $B=M$ of each node of the multiplex networks, in which case, the distribution $P(k,B)$ is given by Eq. \eqref{eq:PkB1}.
Here we consider the other extreme case in which $P(k,B)$ corresponds to a multiplex network in which all nodes have the same intra-layer degree $k_0$, and the multi-modal activity distribution $\tilde{P}(B)$:
\bea \label{eq:PkB2}
P(k,B)=\delta_{k,k_0}\tilde{P}(B),
\eea
where 
\bea
\tilde{P}(B) = c_1\delta_{B,B_1} + c_2\delta_{B,B_2}+c_3\delta_{B,B_3},
\eea
and $c_1+c_2+c_3=1 $.  
Also in this scenario, we observe that if the modes of the activity distribution $\tilde{P}(B)$ are well separated, the interdependent percolation may again result in multiple phase transitions. These phase transitions correspond to the decomposition of the multiplex network into sub-multiplexes having nodes with activity lower or equal to a given threshold. 
If all activities are greater than one,  $B_n>1$ for $n=1,2,3,$ the interdependent percolation features up to three discontinuous phase transitions. In this case the phase diagram contains three regions denoted as $\Omega_{01},\Omega_{02}$, and $\Omega_{03}$ indicating that there are zero continuous and one, two, or three discontinuous transition respectively. However, if we set $B_1=1$ the network may also display one continuous and up to two discontinuous phase transitions. Therefore, in this case, as shown in Fig.~\ref{fig.SMbarycentricA}, we have four possible domains.\\

%%% FIGURE 3 %%%
%%% FIGURE 4 %%%
\begin{figure}[t]
\begin{center}
\includegraphics[width=1\columnwidth]{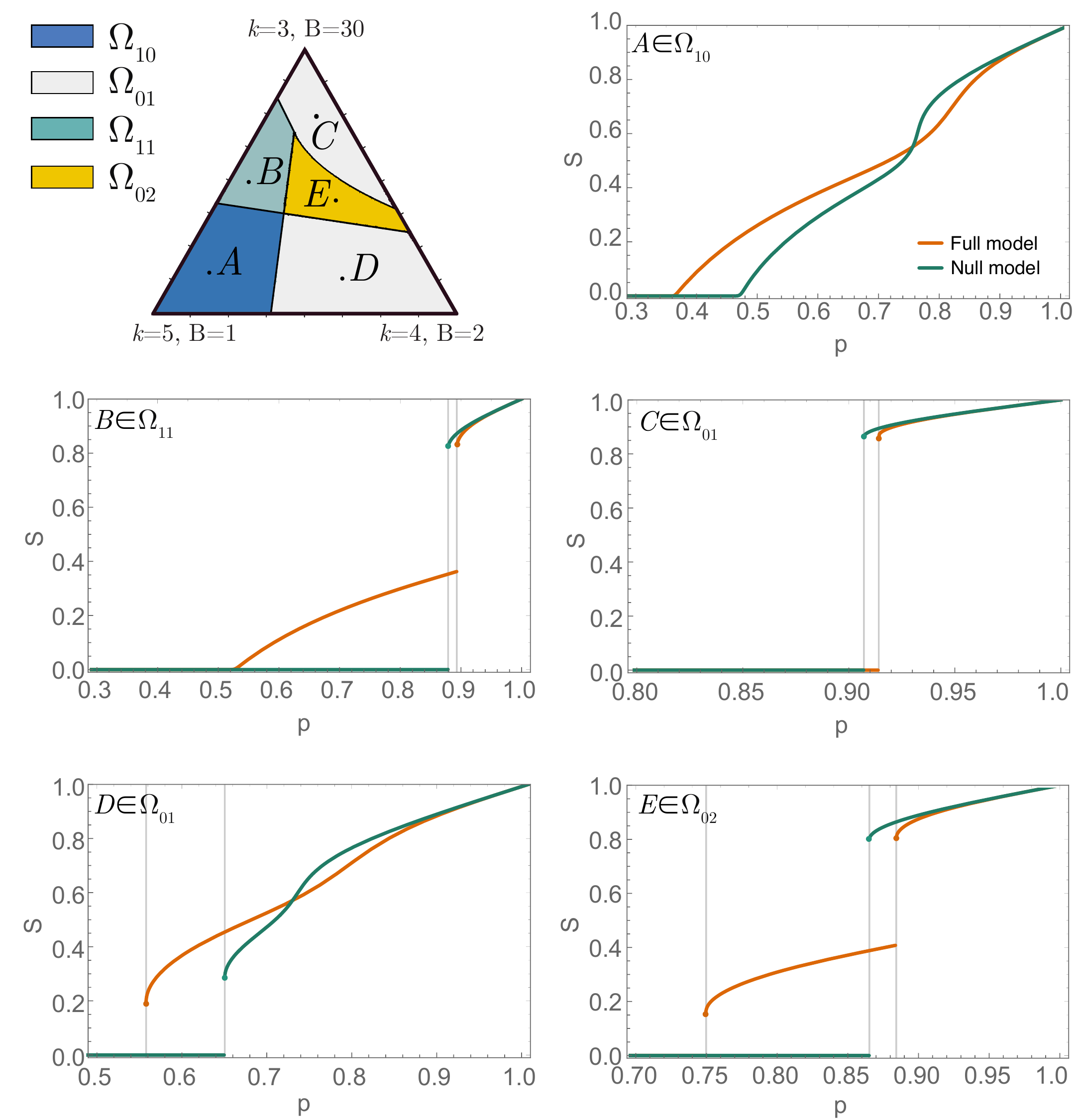}
\caption {{\it Barycentric plot:} the phase diagram for interdependent percolation in the multiplex network with the degree-activity distribution given by $P(k,B)=c_1\delta_{k,5}\delta_{B,1}+c_2\delta_{k,4}\delta_{B,2}+c_3\delta_{k,3}\delta_{B,30}$.
 {\it Panels:}  Comparison of the MCGC size $S$ versus the randomised null model at the  points $A,B,C,D$ and $E$ having coordinates $(c_1,c_2,c_3)$:
 $A = (0.62, 0.10, 0.28),$
 $B = (0.36, 0.08, 0.56),$
 $C = (0.1, 0.2, 0.7),$
$D = (0.32, 0.48, 0.2),$ and
$E = (0.22, 0.30, 0.48).$ 
  \vspace{-0.5cm}}
\label{fig.SMbarycentricB}
\end{center}
\end{figure}

\emph{Correlated intra-layer degree and activity.}
Consider a scenario in which a node's degree and activity correlate:
\bea
P(k,B)=c_1\delta_{k,k_1}\delta_{B,B_1}+c_2\delta_{k,k_2}\delta_{B,B_2}+c_3\delta_{k,k_3}\delta_{B,B_3},
\eea
and thus neither $B_1,B_2,B_3$ nor $k_1,k_2,k_3$ are triplets of identical numbers.  As shown in Figure \ref{fig.SMbarycentricB}, also in this case we see a rich phase diagram including phases $\Omega_{10},\Omega_{01},\Omega_{11}$ and $\Omega_{02}$  having the same definition as above.  To asses the effect of anti-correlations between activity and intra-layer degree, the robustness of  multiplex networks  can  be compared against the null model presented in Appendix B, in which the intra-layer degrees and activity are independent.
The intuition is that anti-correlations between activity and degree will enhance the robustness of the network as the hubs of each layer are less prone to failure than in the null model due to the fact that they are interdependent with a smaller number of layers, we call these nodes super-robust-nodes (SRN). Indeed, in these networks  $p_c$ is typically lower than in the null model. However anti-correlations also reduce the fragility of low degree nodes by associating to them high activity values, we call these nodes super-fragile-nodes (SFN).
Therefore intra-layer degree-activity correlation may lead to both increased and reduced robustness of the multiplex network monitored by measuring the fraction of nodes $S$ in the MCGC depending on the region of the phase diagram and the exact value of percolation parameter $p$ (see Figure \ref{fig.SMbarycentricB}).
For large values of $p$ the percolation process on a multiplex network with degree-activity anti-correlation efficiently dismantles the sub-multiplex formed by the SFN leading to a reduced value of $S$ with respect to the null model. At small  $p$ either the network is totally dismantled (see for instance point $C$ in Figure  \ref{fig.SMbarycentricB}) or the anti-correlated multiplex network is formed by a large percentage SRN leading to an increased value of $S$ and a smaller value of $p_c$.\\

%%% SM FIGURE 5 %%%

%%% SM FIGURE 6 %%%
\begin{figure}
\begin{center}
\includegraphics[width=\columnwidth]{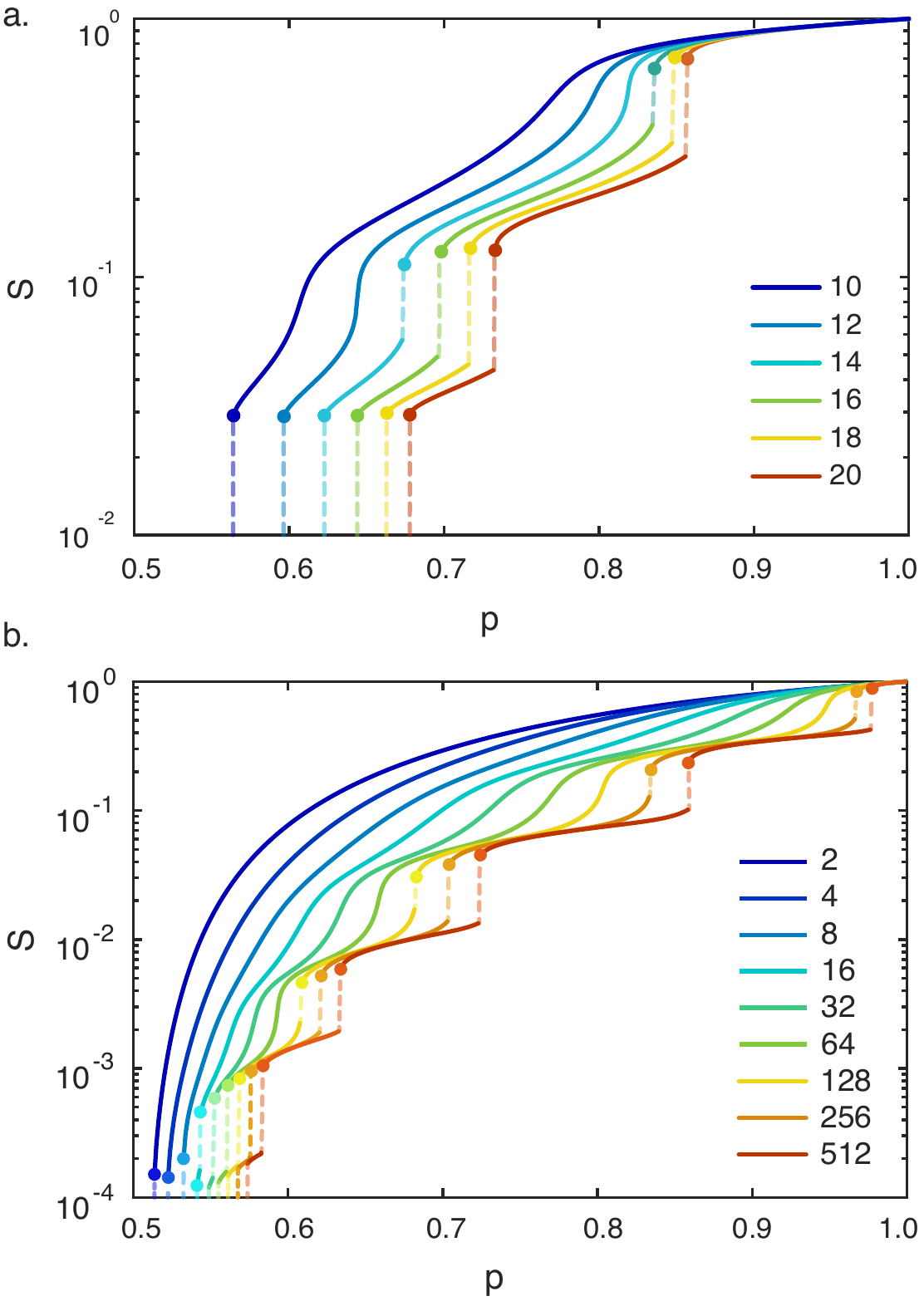}
\caption{
 The effect of number of layers on network robustness.
The size of MCGC $S$ in a network with joint degree-activity distributions given by: 
(a) 
$P(k,B)=\delta_{B,M}\hat{P}(k),$  $\hat{P}(k)=c_1\delta_{k,3}+c_2\delta_{k,9}+c_3\delta_{k,30}$ with $(c_1,c_2,c_3)=(0.75,0.2,0.05)$,
and 
(b)
$P(k,B)=\delta_{B,M}(c_1\delta_{k,3^1}+ c_2\delta_{k,3^2}+ c_3\delta_{k,3^3}+  c_4\delta_{k,3^4}+ c_5\delta_{k,3^5}+c_6\delta_{k,3^6} $ where $(c_1,c_2,c_3,c_4,c_5,c_6)=( 0.5718,0.3049,0.0953,0.0210,0.0057,0.0013)$.
 The numbers of layers $M$ are indicated in the legends.}
\label{fig:Layers}
\end{center}
\end{figure}

\emph{The effect of number of layers.} The above-mentioned examples consider only degree-activity distributions with three modes, however, other modes may also lead to more numerous discontinuous phase transitions reminiscent of cascades in the Barkhausen effect \cite{zapperi1998dynamics}. Figs~\ref{fig:Layers}a,b show that when multiple phase transitions are present, these phase transitions can be removed by reducing the number of layers. In which case the jump discontinuity that corresponds to the largest value of $p_c$ is first to vanish, and eventually, all discontinuous transitions disappear. 
Our results show that interdependent percolation and thus the corresponding interdependent epidemic spreading process defined on multiplex networks with many layers cannot always be effectively approximated with just a few layers.  For instance,  Figs~\ref{fig:Layers}a,b provide examples of a qualitative change of the percolation properties that are associated to a small change in the number of layers (\emph{i.e.} increment or decrement by one) even if the total number of layers is large.  

In this work we have shown evidence of an unexpected outcome of  interdependent percolation on multiplex networks. We showed that multiplex networks with maximum inter-layer degree correlations that were designed to minimize $p_c$  can also display multiple discontinuous phase transitions corresponding to the successive dismantling of sub-multiplexes that span across all the layers of the network. 
Therefore, although setting all degrees of a node's replicas to be identical do yield optimally robust multilayer networks (as this operation minimizes $p_c$), such an increased robustness might come at the expense of introducing a series of discontinuous phase transitions.
This result shows that in the large variety of contexts, when this type of optimization principle might be at work, including engineering, economics and biological networks, constrained or multi-objective optimization \cite{santoro2018pareto,shoval2012evolutionary} should be adopted to avoid multiple discontinuous phase transitions.

Another aspect of this study is the demonstration that even in multiplex networks with large number of layers, removing even one layer may lead to a qualitative change in the percolation behaviour as the number of phase transitions may change. Many real networks do naturally admit a multi-layer representation in which the number of layers is large or even tends to infinity. Examples include airport networks \cite{cardillo,nicosia2015measuring}, trade networks \cite{de2015structural}, social networks \cite{nicosia2015measuring} and  temporal networks \cite{masuda2016guidance,Perra,Valdano} wherein layers represent time. This study shows that it is not always possible to approximate such networks with multiplexes having a smaller number of layers while still preserving the qualitative structure of the phase space.

\begin{acknowledgments}
I.K. acknowledges kind hospitality of the School of Mathematical Sciences at QMUL 
 and the funding from Netherlands Organisation for Scientific Research through the Veni scheme, project number 639.071.511.
\end{acknowledgments}

\appendix
\section*{Appendix A: Percolation in interdependent multiplex networks}

The MCGC can be defined as the subset of nodes in which each pair is connected by a path (internal to the MCGC) in each layer \cite{Baxter}.
 In a multiplex network with joint degree-activity distribution $P(k,B)$ the fraction of nodes $S$ in the MCGC is given by \cite{bianconi2018multilayer,cellai2016}:
\bea\label{eqa:S}
S&=&p\sum_{k,B>0}P(k,B)[1-(1-s)^k]^{B},
\eea
where $s$, the probability that by following a randomly chosen link one arrives at a node that belongs to the MCGC, satisfies the following implicit equation:
\begin{equation}\label{eqa:Sp}
s=p\sum_{k,B>0}\frac{k}{\avg{k}}P(k,B)[1-(1-s)^{k-1}]\\
\times[1-(1-s)^k]^{B-1}.
\end{equation}
The interdependent percolation can thus be fully characterized by studying the critical behavior of Eq. $(\ref{eqa:S})$ and Eq. $(\ref{eqa:Sp})$.
If for a given value of $p$ there exists multiple solutions $s\in [0,1]$ of Eq.~\eqref{eqa:Sp}, the solution having the largest value provides the probability that by following a randomly chosen link one arrives at a node that belongs to the MCGC.
The solutions of Eqs.~\eqref{eqa:S} and \eqref{eqa:Sp} fully characterise the critical behaviour of the interdependent percolation in the interdependent multiplex networks.
Since the number and type of discontinuities of $S(p)$ and $s(p)$ coincide, it suffices to study the behaviour of $s=s(p)$ to obtain the phase diagram.
Let us denote
 $$Q(s) := \sum_{k>0}\frac{k}{\avg{k}}P(k,B)[1-(1-s)^{k-1}][1-(1-s)^k]^{B-1},$$
then, as long as there is a giant component, i.e. as long as $s>0$, we may express the inverse function to $s(p)$ from Eq.~\eqref{eqa:Sp} by simply writing:
 \bea\label{eqa:ps}
 p(s)=\frac{s}{Q(s)}.
 \eea
Therefore, if $s(p)$ features a jump discontinuity at $(p_c,s_c)$ then $p(s)$ has a local minimum at this point:
\begin{equation}\label{eqa:cond}
\begin{aligned}
 p'(s_c)& =0,\\
 p''(s_c) &>0.
 \end{aligned}
 \end{equation}  By plugging \eqref{eqa:ps} into conditions \eqref{eqa:cond}  and assuming that $s_c>0$, one obtains a polynomial equation for $s$:
\bea\label{eqa:Q1}
Q(s)-sQ'(s)=0, \text{ and } {Q''(s)}<0.
\eea
Note that not all pairs $(p(s_i),s_i)$ such that $s_i$ satisfy Eq. \eqref{eqa:Q1} are associated with valid discontinuous critical points, but only those that belong to the maximal positive subsequence of $(p_i,s_i)$ that is simultaneously descending with respect to $p$ and $s$. Thus the discontinuous critical points of \eqref{eqa:Sp} are related to roots of polynomial equation \eqref{eqa:Q1}. In the general case, Eq. \eqref{eqa:Q1} is not solvable in radicals and one relies on root-finding algorithms (see supplementary software \cite{software}). Alternatively,  one can view Eq.~\eqref{eqa:Q1} as the eigenvalue problem for the corresponding companion matrix having its size equal to the maximum degree. It is this connection between the positions of phase transitions and the roots of polynomial \eqref{eqa:Q1} that allows performing an efficient parametric study. 
The numerical analysis present in this this paper reveals how the coefficients of polynomial \eqref{eqa:Q1} influence the number of its roots.

The continuous critical point is identified by imposing the continuity condition $s_c=0$, in which case, plugging  Eq. \eqref{eqa:ps}  into Eq.~\eqref{eqa:cond} gives an explicit expression:
\bea\label{eqa:Q2}
p_c=\frac{1}{Q'(0)},
\eea
which is a valid critical point only if $\frac{1}{Q'(0)}<\min\limits_i p(s_i)$, that is when $\frac{1}{Q'(0)}$ is smaller then the smallest discontinuous point.

In order to investigate whether in presence of multiple percolation transitions the multiplex network effectively decomposes into sub-multiplexes we introduce an additional local order parameter $\sigma_{k,B}$ indicating the  fraction of nodes that have degree $k$, activity $B$ and that belong to the MCGC:
\bea
\sigma_{k,B}&=&pP(k,B)[1-(1-S^{\prime})^k]^{B}.
\label{sigma}
\eea
This local order parameter indicates whether the corresponding sub-multiplexes belong to the MCGC.\\

\section*{Appendix B: Randomised multiplex  models}
In order to investigate the roles of degree and activity correlations, we introduce two randomised null models.
In the first model, we remove the dependence between the node degrees and activity by writing the degree-activity distribution in the factorizable form:
\bea\label{eqa:factor}
P_\text{null}(k,B)=\tilde{P}(B)\hat{P}(k),
\eea
where
\bea
\begin{aligned}
\tilde{P}(B)=\sum_{k>0}{P}(k,B), \text{ and } \hat{P}(k)=\sum_{B>0}{P}(k,B).
\end{aligned}
\eea
In this case, the interdependent percolation can be studied by investigating the critical properties of Eqs. \eqref{eqa:S} and \eqref{eqa:Sp} supplied with $P(k,B)=P_\text{null}(k,B)$. Note that in this randomised model, the degrees of all replicas of a given node are still identical.

In the second randomised model we relax the later constrain by considering a multiplex network in which for a given node the degrees of its replicas  are independent random variables with probability mass function $\hat{P}(k)$. In this case, Eqs. \eqref{eqa:S} and \eqref{eqa:Sp} are to be replaced by:
\bea\label{eqa:S_sigma_unc}
\begin{aligned}
S=&\;p\sum_{k>0}\tilde{P}(B)[1-G_0(1-s)]^{B},\\
s=&\;p\sum_{k>0}\tilde{P}(B)[1-G_1(1-s)] [1-G_0(1-s)]^{B-1},
\end{aligned}
\eea
where 
\bea
G_0(x)=\sum\limits_{k>0}\hat{P}(k) x^k, \; G_1(x)=\sum\limits_{k>0}\frac{k}{\avg{k}}\hat{P}(k) x^{k-1}.
\eea
Moreover, the fraction of nodes that have activity $B$ and belong to the MCGC is given by 
\bea
\sigma_{B}=&\;p\tilde{P}(B)[1-G_0(1-s)]^{B}.
\eea

\bibliographystyle{apsrev4-1}
\bibliography{literature}

\begin{thebibliography}{48}%
\makeatletter
\providecommand \@ifxundefined [1]{%
 \@ifx{#1\undefined}
}%
\providecommand \@ifnum [1]{%
 \ifnum #1\expandafter \@firstoftwo
 \else \expandafter \@secondoftwo
 \fi
}%
\providecommand \@ifx [1]{%
 \ifx #1\expandafter \@firstoftwo
 \else \expandafter \@secondoftwo
 \fi
}%
\providecommand \natexlab [1]{#1}%
\providecommand \enquote  [1]{``#1''}%
\providecommand \bibnamefont  [1]{#1}%
\providecommand \bibfnamefont [1]{#1}%
\providecommand \citenamefont [1]{#1}%
\providecommand \href@noop [0]{\@secondoftwo}%
\providecommand \href [0]{\begingroup \@sanitize@url \@href}%
\providecommand \@href[1]{\@@startlink{#1}\@@href}%
\providecommand \@@href[1]{\endgroup#1\@@endlink}%
\providecommand \@sanitize@url [0]{\catcode `\\12\catcode `\$12\catcode
  `\&12\catcode `\#12\catcode `\^12\catcode `\_12\catcode `\%12\relax}%
\providecommand \@@startlink[1]{}%
\providecommand \@@endlink[0]{}%
\providecommand \url  [0]{\begingroup\@sanitize@url \@url }%
\providecommand \@url [1]{\endgroup\@href {#1}{\urlprefix }}%
\providecommand \urlprefix  [0]{URL }%
\providecommand \Eprint [0]{\href }%
\providecommand \doibase [0]{http://dx.doi.org/}%
\providecommand \selectlanguage [0]{\@gobble}%
\providecommand \bibinfo  [0]{\@secondoftwo}%
\providecommand \bibfield  [0]{\@secondoftwo}%
\providecommand \translation [1]{[#1]}%
\providecommand \BibitemOpen [0]{}%
\providecommand \bibitemStop [0]{}%
\providecommand \bibitemNoStop [0]{.\EOS\space}%
\providecommand \EOS [0]{\spacefactor3000\relax}%
\providecommand \BibitemShut  [1]{\csname bibitem#1\endcsname}%
\let\auto@bib@innerbib\@empty
%</preamble>
\bibitem [{\citenamefont {Bianconi}(2018)}]{bianconi2018multilayer}%
  \BibitemOpen
  \bibfield  {author} {\bibinfo {author} {\bibfnamefont {G.}~\bibnamefont
  {Bianconi}},\ }\href@noop {} {\emph {\bibinfo {title} {Multilayer Networks:
  Structure and Function}}}\ (\bibinfo  {publisher} {Oxford University Press,
  Oxford},\ \bibinfo {year} {2018})\BibitemShut {NoStop}%
\bibitem [{\citenamefont {Boccaletti}\ \emph {et~al.}(2014)\citenamefont
  {Boccaletti}, \citenamefont {Bianconi}, \citenamefont {Criado}, \citenamefont
  {Del~Genio}, \citenamefont {G{\'o}mez-Gardenes}, \citenamefont {Romance},
  \citenamefont {Sendina-Nadal}, \citenamefont {Wang},\ and\ \citenamefont
  {Zanin}}]{PhysReport}%
  \BibitemOpen
  \bibfield  {author} {\bibinfo {author} {\bibfnamefont {S.}~\bibnamefont
  {Boccaletti}}, \bibinfo {author} {\bibfnamefont {G.}~\bibnamefont
  {Bianconi}}, \bibinfo {author} {\bibfnamefont {R.}~\bibnamefont {Criado}},
  \bibinfo {author} {\bibfnamefont {C.~I.}\ \bibnamefont {Del~Genio}}, \bibinfo
  {author} {\bibfnamefont {J.}~\bibnamefont {G{\'o}mez-Gardenes}}, \bibinfo
  {author} {\bibfnamefont {M.}~\bibnamefont {Romance}}, \bibinfo {author}
  {\bibfnamefont {I.}~\bibnamefont {Sendina-Nadal}}, \bibinfo {author}
  {\bibfnamefont {Z.}~\bibnamefont {Wang}}, \ and\ \bibinfo {author}
  {\bibfnamefont {M.}~\bibnamefont {Zanin}},\ }\href@noop {} {\bibfield
  {journal} {\bibinfo  {journal} {Physics Reports}\ }\textbf {\bibinfo {volume}
  {544}},\ \bibinfo {pages} {1} (\bibinfo {year} {2014})}\BibitemShut {NoStop}%
\bibitem [{\citenamefont {Kivel{\"a}}\ \emph {et~al.}(2014)\citenamefont
  {Kivel{\"a}}, \citenamefont {Arenas}, \citenamefont {Barthelemy},
  \citenamefont {Gleeson}, \citenamefont {Moreno},\ and\ \citenamefont
  {Porter}}]{Kivela}%
  \BibitemOpen
  \bibfield  {author} {\bibinfo {author} {\bibfnamefont {M.}~\bibnamefont
  {Kivel{\"a}}}, \bibinfo {author} {\bibfnamefont {A.}~\bibnamefont {Arenas}},
  \bibinfo {author} {\bibfnamefont {M.}~\bibnamefont {Barthelemy}}, \bibinfo
  {author} {\bibfnamefont {J.~P.}\ \bibnamefont {Gleeson}}, \bibinfo {author}
  {\bibfnamefont {Y.}~\bibnamefont {Moreno}}, \ and\ \bibinfo {author}
  {\bibfnamefont {M.~A.}\ \bibnamefont {Porter}},\ }\href@noop {} {\bibfield
  {journal} {\bibinfo  {journal} {Journal of complex networks}\ }\textbf
  {\bibinfo {volume} {2}},\ \bibinfo {pages} {203} (\bibinfo {year}
  {2014})}\BibitemShut {NoStop}%
\bibitem [{\citenamefont {Schamboeck}\ \emph {et~al.}(2019)\citenamefont
  {Schamboeck}, \citenamefont {Iedema},\ and\ \citenamefont
  {Kryven}}]{schamboeck2019}%
  \BibitemOpen
  \bibfield  {author} {\bibinfo {author} {\bibfnamefont {V.}~\bibnamefont
  {Schamboeck}}, \bibinfo {author} {\bibfnamefont {P.~D.}\ \bibnamefont
  {Iedema}}, \ and\ \bibinfo {author} {\bibfnamefont {I.}~\bibnamefont
  {Kryven}},\ }\href@noop {} {\bibfield  {journal} {\bibinfo  {journal}
  {Scientific Reports}\ }\textbf {\bibinfo {volume} {9}},\ \bibinfo {pages}
  {2276} (\bibinfo {year} {2019})}\BibitemShut {NoStop}%
\bibitem [{\citenamefont {Kryven}(2018)}]{kryven2018a}%
  \BibitemOpen
  \bibfield  {author} {\bibinfo {author} {\bibfnamefont {I.}~\bibnamefont
  {Kryven}},\ }\href@noop {} {\bibfield  {journal} {\bibinfo  {journal}
  {Journal of Mathematical Chemistry}\ }\textbf {\bibinfo {volume} {56}},\
  \bibinfo {pages} {140} (\bibinfo {year} {2018})}\BibitemShut {NoStop}%
\bibitem [{\citenamefont {Orlova}\ \emph {et~al.}(2018)\citenamefont {Orlova},
  \citenamefont {Kryven},\ and\ \citenamefont {Iedema}}]{orlova2018automated}%
  \BibitemOpen
  \bibfield  {author} {\bibinfo {author} {\bibfnamefont {Y.}~\bibnamefont
  {Orlova}}, \bibinfo {author} {\bibfnamefont {I.}~\bibnamefont {Kryven}}, \
  and\ \bibinfo {author} {\bibfnamefont {P.~D.}\ \bibnamefont {Iedema}},\
  }\href@noop {} {\bibfield  {journal} {\bibinfo  {journal} {Computers \&
  Chemical Engineering}\ }\textbf {\bibinfo {volume} {112}},\ \bibinfo {pages}
  {37} (\bibinfo {year} {2018})}\BibitemShut {NoStop}%
\bibitem [{\citenamefont {Kryven}\ \emph {et~al.}(2016)\citenamefont {Kryven},
  \citenamefont {Duivenvoorden}, \citenamefont {Hermans},\ and\ \citenamefont
  {Iedema}}]{kryven2016}%
  \BibitemOpen
  \bibfield  {author} {\bibinfo {author} {\bibfnamefont {I.}~\bibnamefont
  {Kryven}}, \bibinfo {author} {\bibfnamefont {J.}~\bibnamefont
  {Duivenvoorden}}, \bibinfo {author} {\bibfnamefont {J.}~\bibnamefont
  {Hermans}}, \ and\ \bibinfo {author} {\bibfnamefont {P.~D.}\ \bibnamefont
  {Iedema}},\ }\href@noop {} {\bibfield  {journal} {\bibinfo  {journal}
  {Macromolecular Theory and Simulations}\ }\textbf {\bibinfo {volume} {25}},\
  \bibinfo {pages} {449} (\bibinfo {year} {2016})}\BibitemShut {NoStop}%
\bibitem [{\citenamefont {Kryven}(2016)}]{Kryven2016b}%
  \BibitemOpen
  \bibfield  {author} {\bibinfo {author} {\bibfnamefont {I.}~\bibnamefont
  {Kryven}},\ }\href@noop {} {\bibfield  {journal} {\bibinfo  {journal} {Phys.
  Rev. E}\ }\textbf {\bibinfo {volume} {94}},\ \bibinfo {pages} {012315}
  (\bibinfo {year} {2016})}\BibitemShut {NoStop}%
\bibitem [{\citenamefont {Buldyrev}\ \emph {et~al.}(2010)\citenamefont
  {Buldyrev}, \citenamefont {Parshani}, \citenamefont {Paul}, \citenamefont
  {Stanley},\ and\ \citenamefont {Havlin}}]{Havlin}%
  \BibitemOpen
  \bibfield  {author} {\bibinfo {author} {\bibfnamefont {S.~V.}\ \bibnamefont
  {Buldyrev}}, \bibinfo {author} {\bibfnamefont {R.}~\bibnamefont {Parshani}},
  \bibinfo {author} {\bibfnamefont {G.}~\bibnamefont {Paul}}, \bibinfo {author}
  {\bibfnamefont {H.~E.}\ \bibnamefont {Stanley}}, \ and\ \bibinfo {author}
  {\bibfnamefont {S.}~\bibnamefont {Havlin}},\ }\href@noop {} {\bibfield
  {journal} {\bibinfo  {journal} {Nature}\ }\textbf {\bibinfo {volume} {464}},\
  \bibinfo {pages} {1025} (\bibinfo {year} {2010})}\BibitemShut {NoStop}%
\bibitem [{\citenamefont {Baxter}\ \emph {et~al.}(2012)\citenamefont {Baxter},
  \citenamefont {Dorogovtsev}, \citenamefont {Goltsev},\ and\ \citenamefont
  {Mendes}}]{Baxter}%
  \BibitemOpen
  \bibfield  {author} {\bibinfo {author} {\bibfnamefont {G.~J.}\ \bibnamefont
  {Baxter}}, \bibinfo {author} {\bibfnamefont {S.~N.}\ \bibnamefont
  {Dorogovtsev}}, \bibinfo {author} {\bibfnamefont {A.~V.}\ \bibnamefont
  {Goltsev}}, \ and\ \bibinfo {author} {\bibfnamefont {J.~F.~F.}\ \bibnamefont
  {Mendes}},\ }\href@noop {} {\bibfield  {journal} {\bibinfo  {journal}
  {Physical Review Letters}\ }\textbf {\bibinfo {volume} {109}},\ \bibinfo
  {pages} {248701} (\bibinfo {year} {2012})}\BibitemShut {NoStop}%
\bibitem [{\citenamefont {Son}\ \emph {et~al.}(2012)\citenamefont {Son},
  \citenamefont {Bizhani}, \citenamefont {Christensen}, \citenamefont
  {Grassberger},\ and\ \citenamefont {Paczuski}}]{Son}%
  \BibitemOpen
  \bibfield  {author} {\bibinfo {author} {\bibfnamefont {S.-W.}\ \bibnamefont
  {Son}}, \bibinfo {author} {\bibfnamefont {G.}~\bibnamefont {Bizhani}},
  \bibinfo {author} {\bibfnamefont {C.}~\bibnamefont {Christensen}}, \bibinfo
  {author} {\bibfnamefont {P.}~\bibnamefont {Grassberger}}, \ and\ \bibinfo
  {author} {\bibfnamefont {M.}~\bibnamefont {Paczuski}},\ }\href@noop {}
  {\bibfield  {journal} {\bibinfo  {journal} {Europhysics Letters}\ }\textbf
  {\bibinfo {volume} {97}},\ \bibinfo {pages} {16006} (\bibinfo {year}
  {2012})}\BibitemShut {NoStop}%
\bibitem [{\citenamefont {Parshani}\ \emph {et~al.}(2010)\citenamefont
  {Parshani}, \citenamefont {Buldyrev},\ and\ \citenamefont
  {Havlin}}]{Havlin2}%
  \BibitemOpen
  \bibfield  {author} {\bibinfo {author} {\bibfnamefont {R.}~\bibnamefont
  {Parshani}}, \bibinfo {author} {\bibfnamefont {S.~V.}\ \bibnamefont
  {Buldyrev}}, \ and\ \bibinfo {author} {\bibfnamefont {S.}~\bibnamefont
  {Havlin}},\ }\href@noop {} {\bibfield  {journal} {\bibinfo  {journal}
  {Physical Review Letters}\ }\textbf {\bibinfo {volume} {105}},\ \bibinfo
  {pages} {048701} (\bibinfo {year} {2010})}\BibitemShut {NoStop}%
\bibitem [{\citenamefont {Lee}\ \emph {et~al.}(2016)\citenamefont {Lee},
  \citenamefont {Choi}, \citenamefont {Stippinger}, \citenamefont
  {Kert{\'e}sz},\ and\ \citenamefont {Kahng}}]{Kahng1}%
  \BibitemOpen
  \bibfield  {author} {\bibinfo {author} {\bibfnamefont {D.}~\bibnamefont
  {Lee}}, \bibinfo {author} {\bibfnamefont {S.}~\bibnamefont {Choi}}, \bibinfo
  {author} {\bibfnamefont {M.}~\bibnamefont {Stippinger}}, \bibinfo {author}
  {\bibfnamefont {J.}~\bibnamefont {Kert{\'e}sz}}, \ and\ \bibinfo {author}
  {\bibfnamefont {B.}~\bibnamefont {Kahng}},\ }\href@noop {} {\bibfield
  {journal} {\bibinfo  {journal} {Physical Review E}\ }\textbf {\bibinfo
  {volume} {93}},\ \bibinfo {pages} {042109} (\bibinfo {year}
  {2016})}\BibitemShut {NoStop}%
\bibitem [{\citenamefont {Lee}\ \emph {et~al.}(2017)\citenamefont {Lee},
  \citenamefont {Choi}, \citenamefont {K{\'e}rtesz},\ and\ \citenamefont
  {Kahng}}]{Kahng2}%
  \BibitemOpen
  \bibfield  {author} {\bibinfo {author} {\bibfnamefont {D.}~\bibnamefont
  {Lee}}, \bibinfo {author} {\bibfnamefont {W.}~\bibnamefont {Choi}}, \bibinfo
  {author} {\bibfnamefont {J.}~\bibnamefont {K{\'e}rtesz}}, \ and\ \bibinfo
  {author} {\bibfnamefont {B.}~\bibnamefont {Kahng}},\ }\href@noop {}
  {\bibfield  {journal} {\bibinfo  {journal} {Scientific Reports}\ }\textbf
  {\bibinfo {volume} {7}},\ \bibinfo {pages} {5723} (\bibinfo {year}
  {2017})}\BibitemShut {NoStop}%
\bibitem [{\citenamefont {Rapisardi}\ \emph {et~al.}(2019)\citenamefont
  {Rapisardi}, \citenamefont {Arenas}, \citenamefont {Caldarelli},\ and\
  \citenamefont {Cimini}}]{rapisardi2019fragility}%
  \BibitemOpen
  \bibfield  {author} {\bibinfo {author} {\bibfnamefont {G.}~\bibnamefont
  {Rapisardi}}, \bibinfo {author} {\bibfnamefont {A.}~\bibnamefont {Arenas}},
  \bibinfo {author} {\bibfnamefont {G.}~\bibnamefont {Caldarelli}}, \ and\
  \bibinfo {author} {\bibfnamefont {G.}~\bibnamefont {Cimini}},\ }\href@noop {}
  {\bibfield  {journal} {\bibinfo  {journal} {Physical Review E}\ }\textbf
  {\bibinfo {volume} {99}},\ \bibinfo {pages} {042302} (\bibinfo {year}
  {2019})}\BibitemShut {NoStop}%
\bibitem [{\citenamefont {Min}\ \emph {et~al.}(2014)\citenamefont {Min},
  \citenamefont {Yi}, \citenamefont {Lee},\ and\ \citenamefont {Goh}}]{Goh}%
  \BibitemOpen
  \bibfield  {author} {\bibinfo {author} {\bibfnamefont {B.}~\bibnamefont
  {Min}}, \bibinfo {author} {\bibfnamefont {S.~D.}\ \bibnamefont {Yi}},
  \bibinfo {author} {\bibfnamefont {K.-M.}\ \bibnamefont {Lee}}, \ and\
  \bibinfo {author} {\bibfnamefont {K.-I.}\ \bibnamefont {Goh}},\ }\href@noop
  {} {\bibfield  {journal} {\bibinfo  {journal} {Physical Review E}\ }\textbf
  {\bibinfo {volume} {89}},\ \bibinfo {pages} {042811} (\bibinfo {year}
  {2014})}\BibitemShut {NoStop}%
\bibitem [{\citenamefont {Reis}\ \emph {et~al.}(2014)\citenamefont {Reis},
  \citenamefont {Hu}, \citenamefont {Babino}, \citenamefont {Andrade~Jr},
  \citenamefont {Canals}, \citenamefont {Sigman},\ and\ \citenamefont
  {Makse}}]{Makse}%
  \BibitemOpen
  \bibfield  {author} {\bibinfo {author} {\bibfnamefont {S.~D.}\ \bibnamefont
  {Reis}}, \bibinfo {author} {\bibfnamefont {Y.}~\bibnamefont {Hu}}, \bibinfo
  {author} {\bibfnamefont {A.}~\bibnamefont {Babino}}, \bibinfo {author}
  {\bibfnamefont {J.~S.}\ \bibnamefont {Andrade~Jr}}, \bibinfo {author}
  {\bibfnamefont {S.}~\bibnamefont {Canals}}, \bibinfo {author} {\bibfnamefont
  {M.}~\bibnamefont {Sigman}}, \ and\ \bibinfo {author} {\bibfnamefont {H.~A.}\
  \bibnamefont {Makse}},\ }\href@noop {} {\bibfield  {journal} {\bibinfo
  {journal} {Nature Physics}\ }\textbf {\bibinfo {volume} {10}},\ \bibinfo
  {pages} {762} (\bibinfo {year} {2014})}\BibitemShut {NoStop}%
\bibitem [{\citenamefont {Radicchi}\ and\ \citenamefont
  {Bianconi}(2017)}]{Redundant}%
  \BibitemOpen
  \bibfield  {author} {\bibinfo {author} {\bibfnamefont {F.}~\bibnamefont
  {Radicchi}}\ and\ \bibinfo {author} {\bibfnamefont {G.}~\bibnamefont
  {Bianconi}},\ }\href@noop {} {\bibfield  {journal} {\bibinfo  {journal}
  {Physical Review X}\ }\textbf {\bibinfo {volume} {7}},\ \bibinfo {pages}
  {011013} (\bibinfo {year} {2017})}\BibitemShut {NoStop}%
\bibitem [{\citenamefont {Cellai}\ \emph {et~al.}(2013)\citenamefont {Cellai},
  \citenamefont {L{\'o}pez}, \citenamefont {Zhou}, \citenamefont {Gleeson},\
  and\ \citenamefont {Bianconi}}]{cellai2013}%
  \BibitemOpen
  \bibfield  {author} {\bibinfo {author} {\bibfnamefont {D.}~\bibnamefont
  {Cellai}}, \bibinfo {author} {\bibfnamefont {E.}~\bibnamefont {L{\'o}pez}},
  \bibinfo {author} {\bibfnamefont {J.}~\bibnamefont {Zhou}}, \bibinfo {author}
  {\bibfnamefont {J.~P.}\ \bibnamefont {Gleeson}}, \ and\ \bibinfo {author}
  {\bibfnamefont {G.}~\bibnamefont {Bianconi}},\ }\href@noop {} {\bibfield
  {journal} {\bibinfo  {journal} {Physical Review E}\ }\textbf {\bibinfo
  {volume} {88}},\ \bibinfo {pages} {052811} (\bibinfo {year}
  {2013})}\BibitemShut {NoStop}%
\bibitem [{\citenamefont {Min}\ \emph {et~al.}(2015)\citenamefont {Min},
  \citenamefont {Lee}, \citenamefont {Lee},\ and\ \citenamefont
  {Goh}}]{Goh_viability}%
  \BibitemOpen
  \bibfield  {author} {\bibinfo {author} {\bibfnamefont {B.}~\bibnamefont
  {Min}}, \bibinfo {author} {\bibfnamefont {S.}~\bibnamefont {Lee}}, \bibinfo
  {author} {\bibfnamefont {K.-M.}\ \bibnamefont {Lee}}, \ and\ \bibinfo
  {author} {\bibfnamefont {K.-I.}\ \bibnamefont {Goh}},\ }\href@noop {}
  {\bibfield  {journal} {\bibinfo  {journal} {Chaos, Solitons \& Fractals}\
  }\textbf {\bibinfo {volume} {72}},\ \bibinfo {pages} {49} (\bibinfo {year}
  {2015})}\BibitemShut {NoStop}%
\bibitem [{\citenamefont {Cellai}\ \emph {et~al.}(2016)\citenamefont {Cellai},
  \citenamefont {Dorogovtsev},\ and\ \citenamefont {Bianconi}}]{cellai2016b}%
  \BibitemOpen
  \bibfield  {author} {\bibinfo {author} {\bibfnamefont {D.}~\bibnamefont
  {Cellai}}, \bibinfo {author} {\bibfnamefont {S.~N.}\ \bibnamefont
  {Dorogovtsev}}, \ and\ \bibinfo {author} {\bibfnamefont {G.}~\bibnamefont
  {Bianconi}},\ }\href@noop {} {\bibfield  {journal} {\bibinfo  {journal}
  {Physical Review E}\ }\textbf {\bibinfo {volume} {94}},\ \bibinfo {pages}
  {032301} (\bibinfo {year} {2016})}\BibitemShut {NoStop}%
\bibitem [{\citenamefont {Aldous}\ and\ \citenamefont
  {Barthelemy}(2019)}]{aldous2019optimal}%
  \BibitemOpen
  \bibfield  {author} {\bibinfo {author} {\bibfnamefont {D.}~\bibnamefont
  {Aldous}}\ and\ \bibinfo {author} {\bibfnamefont {M.}~\bibnamefont
  {Barthelemy}},\ }\href {\doibase 10.1103/PhysRevE.99.052303} {\bibfield
  {journal} {\bibinfo  {journal} {Phys. Rev. E}\ }\textbf {\bibinfo {volume}
  {99}},\ \bibinfo {pages} {052303} (\bibinfo {year} {2019})}\BibitemShut
  {NoStop}%
\bibitem [{\citenamefont {Valente}\ \emph {et~al.}(2004)\citenamefont
  {Valente}, \citenamefont {Sarkar},\ and\ \citenamefont
  {Stone}}]{valente2004}%
  \BibitemOpen
  \bibfield  {author} {\bibinfo {author} {\bibfnamefont {A.~X. C.~N.}\
  \bibnamefont {Valente}}, \bibinfo {author} {\bibfnamefont {A.}~\bibnamefont
  {Sarkar}}, \ and\ \bibinfo {author} {\bibfnamefont {H.~A.}\ \bibnamefont
  {Stone}},\ }\href@noop {} {\bibfield  {journal} {\bibinfo  {journal}
  {Physical Review Letters}\ }\textbf {\bibinfo {volume} {92}},\ \bibinfo
  {pages} {118702} (\bibinfo {year} {2004})}\BibitemShut {NoStop}%
\bibitem [{\citenamefont {Paul}\ \emph {et~al.}(2004)\citenamefont {Paul},
  \citenamefont {Tanizawa}, \citenamefont {Havlin},\ and\ \citenamefont
  {Stanley}}]{paul2004}%
  \BibitemOpen
  \bibfield  {author} {\bibinfo {author} {\bibfnamefont {G.}~\bibnamefont
  {Paul}}, \bibinfo {author} {\bibfnamefont {T.}~\bibnamefont {Tanizawa}},
  \bibinfo {author} {\bibfnamefont {S.}~\bibnamefont {Havlin}}, \ and\ \bibinfo
  {author} {\bibfnamefont {H.~E.}\ \bibnamefont {Stanley}},\ }\href@noop {}
  {\bibfield  {journal} {\bibinfo  {journal} {The European Physical Journal B}\
  }\textbf {\bibinfo {volume} {38}},\ \bibinfo {pages} {187} (\bibinfo {year}
  {2004})}\BibitemShut {NoStop}%
\bibitem [{\citenamefont {Paul}\ \emph {et~al.}(2005)\citenamefont {Paul},
  \citenamefont {Sreenivasan},\ and\ \citenamefont {Stanley}}]{paul2005}%
  \BibitemOpen
  \bibfield  {author} {\bibinfo {author} {\bibfnamefont {G.}~\bibnamefont
  {Paul}}, \bibinfo {author} {\bibfnamefont {S.}~\bibnamefont {Sreenivasan}}, \
  and\ \bibinfo {author} {\bibfnamefont {H.~E.}\ \bibnamefont {Stanley}},\
  }\href@noop {} {\bibfield  {journal} {\bibinfo  {journal} {Physical Review
  E}\ }\textbf {\bibinfo {volume} {72}},\ \bibinfo {pages} {056130} (\bibinfo
  {year} {2005})}\BibitemShut {NoStop}%
\bibitem [{\citenamefont {Tanizawa}\ \emph {et~al.}(2005)\citenamefont
  {Tanizawa}, \citenamefont {Paul}, \citenamefont {Cohen}, \citenamefont
  {Havlin},\ and\ \citenamefont {Stanley}}]{tanizawa2005}%
  \BibitemOpen
  \bibfield  {author} {\bibinfo {author} {\bibfnamefont {T.}~\bibnamefont
  {Tanizawa}}, \bibinfo {author} {\bibfnamefont {G.}~\bibnamefont {Paul}},
  \bibinfo {author} {\bibfnamefont {R.}~\bibnamefont {Cohen}}, \bibinfo
  {author} {\bibfnamefont {S.}~\bibnamefont {Havlin}}, \ and\ \bibinfo {author}
  {\bibfnamefont {H.~E.}\ \bibnamefont {Stanley}},\ }\href@noop {} {\bibfield
  {journal} {\bibinfo  {journal} {Physical review E}\ }\textbf {\bibinfo
  {volume} {71}},\ \bibinfo {pages} {047101} (\bibinfo {year}
  {2005})}\BibitemShut {NoStop}%
\bibitem [{\citenamefont {Tanizawa}\ \emph {et~al.}(2006)\citenamefont
  {Tanizawa}, \citenamefont {Paul}, \citenamefont {Havlin},\ and\ \citenamefont
  {Stanley}}]{tanizawa2006}%
  \BibitemOpen
  \bibfield  {author} {\bibinfo {author} {\bibfnamefont {T.}~\bibnamefont
  {Tanizawa}}, \bibinfo {author} {\bibfnamefont {G.}~\bibnamefont {Paul}},
  \bibinfo {author} {\bibfnamefont {S.}~\bibnamefont {Havlin}}, \ and\ \bibinfo
  {author} {\bibfnamefont {H.~E.}\ \bibnamefont {Stanley}},\ }\href@noop {}
  {\bibfield  {journal} {\bibinfo  {journal} {Physical Review E}\ }\textbf
  {\bibinfo {volume} {74}},\ \bibinfo {pages} {016125} (\bibinfo {year}
  {2006})}\BibitemShut {NoStop}%
\bibitem [{\citenamefont {Guimer\'a}\ \emph {et~al.}(2002)\citenamefont
  {Guimer\'a}, \citenamefont {D\'iaz-Guilera}, \citenamefont {Vega-Redondo},
  \citenamefont {Cabrales},\ and\ \citenamefont {Arenas}}]{guimera2002optimal}%
  \BibitemOpen
  \bibfield  {author} {\bibinfo {author} {\bibfnamefont {R.}~\bibnamefont
  {Guimer\'a}}, \bibinfo {author} {\bibfnamefont {A.}~\bibnamefont
  {D\'iaz-Guilera}}, \bibinfo {author} {\bibfnamefont {F.}~\bibnamefont
  {Vega-Redondo}}, \bibinfo {author} {\bibfnamefont {A.}~\bibnamefont
  {Cabrales}}, \ and\ \bibinfo {author} {\bibfnamefont {A.}~\bibnamefont
  {Arenas}},\ }\href@noop {} {\bibfield  {journal} {\bibinfo  {journal}
  {Physical Review Letters}\ }\textbf {\bibinfo {volume} {89}},\ \bibinfo
  {pages} {248701} (\bibinfo {year} {2002})}\BibitemShut {NoStop}%
\bibitem [{\citenamefont {Ferrer~i Cancho}\ and\ \citenamefont
  {Sol\'e}(2003)}]{sole2003optimization}%
  \BibitemOpen
  \bibfield  {author} {\bibinfo {author} {\bibfnamefont {R.}~\bibnamefont
  {Ferrer~i Cancho}}\ and\ \bibinfo {author} {\bibfnamefont {R.~V.}\
  \bibnamefont {Sol\'e}},\ }in\ \href@noop {} {\emph {\bibinfo {booktitle}
  {Statistical mechanics of complex networks}}}\ (\bibinfo  {publisher}
  {Springer},\ \bibinfo {year} {2003})\ pp.\ \bibinfo {pages}
  {114--126}\BibitemShut {NoStop}%
\bibitem [{\citenamefont {Thai}\ and\ \citenamefont
  {Pardalos}(2011)}]{thai2011handbook}%
  \BibitemOpen
  \bibfield  {author} {\bibinfo {author} {\bibfnamefont {M.~T.}\ \bibnamefont
  {Thai}}\ and\ \bibinfo {author} {\bibfnamefont {P.~M.}\ \bibnamefont
  {Pardalos}},\ }\href@noop {} {\emph {\bibinfo {title} {Handbook of
  optimization in complex networks: theory and applications}}},\ Vol.~\bibinfo
  {volume} {57}\ (\bibinfo  {publisher} {Springer science \& business media},\
  \bibinfo {year} {2011})\BibitemShut {NoStop}%
\bibitem [{\citenamefont {Bianconi}\ and\ \citenamefont
  {Dorogovtsev}(2014)}]{bianconi2014}%
  \BibitemOpen
  \bibfield  {author} {\bibinfo {author} {\bibfnamefont {G.}~\bibnamefont
  {Bianconi}}\ and\ \bibinfo {author} {\bibfnamefont {S.~N.}\ \bibnamefont
  {Dorogovtsev}},\ }\href@noop {} {\bibfield  {journal} {\bibinfo  {journal}
  {Physical Review E}\ }\textbf {\bibinfo {volume} {89}},\ \bibinfo {pages}
  {062814} (\bibinfo {year} {2014})}\BibitemShut {NoStop}%
\bibitem [{\citenamefont {{Colomer-de-Sim\'on}}\ and\ \citenamefont
  {Bogu\~n\'a}(2014)}]{colomer2014}%
  \BibitemOpen
  \bibfield  {author} {\bibinfo {author} {\bibfnamefont {P.}~\bibnamefont
  {{Colomer-de-Sim\'on}}}\ and\ \bibinfo {author} {\bibfnamefont
  {M.}~\bibnamefont {Bogu\~n\'a}},\ }\href@noop {} {\bibfield  {journal}
  {\bibinfo  {journal} {Physical Review X}\ }\textbf {\bibinfo {volume} {4}},\
  \bibinfo {pages} {041020} (\bibinfo {year} {2014})}\BibitemShut {NoStop}%
\bibitem [{\citenamefont {Kryven}(2017)}]{kryven2017}%
  \BibitemOpen
  \bibfield  {author} {\bibinfo {author} {\bibfnamefont {I.}~\bibnamefont
  {Kryven}},\ }\href@noop {} {\bibfield  {journal} {\bibinfo  {journal}
  {Physical Review E}\ }\textbf {\bibinfo {volume} {96}},\ \bibinfo {pages}
  {052304} (\bibinfo {year} {2017})}\BibitemShut {NoStop}%
\bibitem [{\citenamefont {Baxter}\ \emph {et~al.}(2016)\citenamefont {Baxter},
  \citenamefont {Bianconi}, \citenamefont {da~Costa}, \citenamefont
  {Dorogovtsev},\ and\ \citenamefont {Mendes}}]{costa}%
  \BibitemOpen
  \bibfield  {author} {\bibinfo {author} {\bibfnamefont {G.~J.}\ \bibnamefont
  {Baxter}}, \bibinfo {author} {\bibfnamefont {G.}~\bibnamefont {Bianconi}},
  \bibinfo {author} {\bibfnamefont {R.~A.}\ \bibnamefont {da~Costa}}, \bibinfo
  {author} {\bibfnamefont {S.~N.}\ \bibnamefont {Dorogovtsev}}, \ and\ \bibinfo
  {author} {\bibfnamefont {J.~F.~F.}\ \bibnamefont {Mendes}},\ }\href@noop {}
  {\bibfield  {journal} {\bibinfo  {journal} {Physical Review E}\ }\textbf
  {\bibinfo {volume} {94}},\ \bibinfo {pages} {012303} (\bibinfo {year}
  {2016})}\BibitemShut {NoStop}%
\bibitem [{\citenamefont {Kryven}(2019)}]{kryven2019}%
  \BibitemOpen
  \bibfield  {author} {\bibinfo {author} {\bibfnamefont {I.}~\bibnamefont
  {Kryven}},\ }\href@noop {} {\bibfield  {journal} {\bibinfo  {journal} {Nature
  communications}\ }\textbf {\bibinfo {volume} {10}},\ \bibinfo {pages} {404}
  (\bibinfo {year} {2019})}\BibitemShut {NoStop}%
\bibitem [{\citenamefont {De~Domenico}\ \emph {et~al.}(2016)\citenamefont
  {De~Domenico}, \citenamefont {Granell}, \citenamefont {Porter},\ and\
  \citenamefont {Arenas}}]{Alex}%
  \BibitemOpen
  \bibfield  {author} {\bibinfo {author} {\bibfnamefont {M.}~\bibnamefont
  {De~Domenico}}, \bibinfo {author} {\bibfnamefont {C.}~\bibnamefont
  {Granell}}, \bibinfo {author} {\bibfnamefont {M.~A.}\ \bibnamefont {Porter}},
  \ and\ \bibinfo {author} {\bibfnamefont {A.}~\bibnamefont {Arenas}},\
  }\href@noop {} {\bibfield  {journal} {\bibinfo  {journal} {Nature Physics}\
  }\textbf {\bibinfo {volume} {12}},\ \bibinfo {pages} {901} (\bibinfo {year}
  {2016})}\BibitemShut {NoStop}%
\bibitem [{\citenamefont {Wang}\ \emph {et~al.}(2019)\citenamefont {Wang},
  \citenamefont {Liu}, \citenamefont {Liang}, \citenamefont {Hu},\ and\
  \citenamefont {Zhou}}]{wang2019}%
  \BibitemOpen
  \bibfield  {author} {\bibinfo {author} {\bibfnamefont {W.}~\bibnamefont
  {Wang}}, \bibinfo {author} {\bibfnamefont {Q.-H.}\ \bibnamefont {Liu}},
  \bibinfo {author} {\bibfnamefont {J.}~\bibnamefont {Liang}}, \bibinfo
  {author} {\bibfnamefont {Y.}~\bibnamefont {Hu}}, \ and\ \bibinfo {author}
  {\bibfnamefont {T.}~\bibnamefont {Zhou}},\ }\href@noop {} {\bibfield
  {journal} {\bibinfo  {journal} {arXiv preprint arXiv:1901.02125}\ } (\bibinfo
  {year} {2019})}\BibitemShut {NoStop}%
\bibitem [{\citenamefont {Masuda}\ and\ \citenamefont
  {Lambiotte}(2016)}]{masuda2016guidance}%
  \BibitemOpen
  \bibfield  {author} {\bibinfo {author} {\bibfnamefont {N.}~\bibnamefont
  {Masuda}}\ and\ \bibinfo {author} {\bibfnamefont {R.}~\bibnamefont
  {Lambiotte}},\ }\href@noop {} {\emph {\bibinfo {title} {A guide to temporal
  networks}}}\ (\bibinfo  {publisher} {World Scientific},\ \bibinfo {year}
  {2016})\BibitemShut {NoStop}%
\bibitem [{\citenamefont {Zapperi}\ \emph {et~al.}(1998)\citenamefont
  {Zapperi}, \citenamefont {Cizeau}, \citenamefont {Durin},\ and\ \citenamefont
  {Stanley}}]{zapperi1998dynamics}%
  \BibitemOpen
  \bibfield  {author} {\bibinfo {author} {\bibfnamefont {S.}~\bibnamefont
  {Zapperi}}, \bibinfo {author} {\bibfnamefont {P.}~\bibnamefont {Cizeau}},
  \bibinfo {author} {\bibfnamefont {G.}~\bibnamefont {Durin}}, \ and\ \bibinfo
  {author} {\bibfnamefont {H.~E.}\ \bibnamefont {Stanley}},\ }\href@noop {}
  {\bibfield  {journal} {\bibinfo  {journal} {Physical Review B}\ }\textbf
  {\bibinfo {volume} {58}},\ \bibinfo {pages} {6353} (\bibinfo {year}
  {1998})}\BibitemShut {NoStop}%
\bibitem [{\citenamefont {Santoro}\ \emph {et~al.}(2018)\citenamefont
  {Santoro}, \citenamefont {Latora}, \citenamefont {Nicosia},\ and\
  \citenamefont {Nicosia}}]{santoro2018pareto}%
  \BibitemOpen
  \bibfield  {author} {\bibinfo {author} {\bibfnamefont {A.}~\bibnamefont
  {Santoro}}, \bibinfo {author} {\bibfnamefont {V.}~\bibnamefont {Latora}},
  \bibinfo {author} {\bibfnamefont {G.}~\bibnamefont {Nicosia}}, \ and\
  \bibinfo {author} {\bibfnamefont {V.}~\bibnamefont {Nicosia}},\ }\href@noop
  {} {\bibfield  {journal} {\bibinfo  {journal} {Physical Review Letters}\
  }\textbf {\bibinfo {volume} {121}},\ \bibinfo {pages} {128302} (\bibinfo
  {year} {2018})}\BibitemShut {NoStop}%
\bibitem [{\citenamefont {Shoval}\ \emph {et~al.}(2012)\citenamefont {Shoval},
  \citenamefont {Sheftel}, \citenamefont {Shinar}, \citenamefont {Hart},
  \citenamefont {Ramote}, \citenamefont {Mayo}, \citenamefont {Dekel},
  \citenamefont {Kavanagh},\ and\ \citenamefont
  {Alon}}]{shoval2012evolutionary}%
  \BibitemOpen
  \bibfield  {author} {\bibinfo {author} {\bibfnamefont {O.}~\bibnamefont
  {Shoval}}, \bibinfo {author} {\bibfnamefont {H.}~\bibnamefont {Sheftel}},
  \bibinfo {author} {\bibfnamefont {G.}~\bibnamefont {Shinar}}, \bibinfo
  {author} {\bibfnamefont {Y.}~\bibnamefont {Hart}}, \bibinfo {author}
  {\bibfnamefont {O.}~\bibnamefont {Ramote}}, \bibinfo {author} {\bibfnamefont
  {A.}~\bibnamefont {Mayo}}, \bibinfo {author} {\bibfnamefont {E.}~\bibnamefont
  {Dekel}}, \bibinfo {author} {\bibfnamefont {K.}~\bibnamefont {Kavanagh}}, \
  and\ \bibinfo {author} {\bibfnamefont {U.}~\bibnamefont {Alon}},\ }\href@noop
  {} {\bibfield  {journal} {\bibinfo  {journal} {Science}\ }\textbf {\bibinfo
  {volume} {336}},\ \bibinfo {pages} {1157} (\bibinfo {year}
  {2012})}\BibitemShut {NoStop}%
\bibitem [{\citenamefont {Cardillo}\ \emph {et~al.}(2013)\citenamefont
  {Cardillo}, \citenamefont {G{\'o}mez-Gardenes}, \citenamefont {Zanin},
  \citenamefont {Romance}, \citenamefont {Papo}, \citenamefont {Del~Pozo},\
  and\ \citenamefont {Boccaletti}}]{cardillo}%
  \BibitemOpen
  \bibfield  {author} {\bibinfo {author} {\bibfnamefont {A.}~\bibnamefont
  {Cardillo}}, \bibinfo {author} {\bibfnamefont {J.}~\bibnamefont
  {G{\'o}mez-Gardenes}}, \bibinfo {author} {\bibfnamefont {M.}~\bibnamefont
  {Zanin}}, \bibinfo {author} {\bibfnamefont {M.}~\bibnamefont {Romance}},
  \bibinfo {author} {\bibfnamefont {D.}~\bibnamefont {Papo}}, \bibinfo {author}
  {\bibfnamefont {F.}~\bibnamefont {Del~Pozo}}, \ and\ \bibinfo {author}
  {\bibfnamefont {S.}~\bibnamefont {Boccaletti}},\ }\href@noop {} {\bibfield
  {journal} {\bibinfo  {journal} {Scientific Reports}\ }\textbf {\bibinfo
  {volume} {3}},\ \bibinfo {pages} {1344} (\bibinfo {year} {2013})}\BibitemShut
  {NoStop}%
\bibitem [{\citenamefont {Nicosia}\ and\ \citenamefont
  {Latora}(2015)}]{nicosia2015measuring}%
  \BibitemOpen
  \bibfield  {author} {\bibinfo {author} {\bibfnamefont {V.}~\bibnamefont
  {Nicosia}}\ and\ \bibinfo {author} {\bibfnamefont {V.}~\bibnamefont
  {Latora}},\ }\href@noop {} {\bibfield  {journal} {\bibinfo  {journal}
  {Physical Review E}\ }\textbf {\bibinfo {volume} {92}},\ \bibinfo {pages}
  {032805} (\bibinfo {year} {2015})}\BibitemShut {NoStop}%
\bibitem [{\citenamefont {De~Domenico}\ \emph {et~al.}(2015)\citenamefont
  {De~Domenico}, \citenamefont {Nicosia}, \citenamefont {Arenas},\ and\
  \citenamefont {Latora}}]{de2015structural}%
  \BibitemOpen
  \bibfield  {author} {\bibinfo {author} {\bibfnamefont {M.}~\bibnamefont
  {De~Domenico}}, \bibinfo {author} {\bibfnamefont {V.}~\bibnamefont
  {Nicosia}}, \bibinfo {author} {\bibfnamefont {A.}~\bibnamefont {Arenas}}, \
  and\ \bibinfo {author} {\bibfnamefont {V.}~\bibnamefont {Latora}},\
  }\href@noop {} {\bibfield  {journal} {\bibinfo  {journal} {Nature
  communications}\ }\textbf {\bibinfo {volume} {6}},\ \bibinfo {pages} {6864}
  (\bibinfo {year} {2015})}\BibitemShut {NoStop}%
\bibitem [{\citenamefont {Perra}\ \emph {et~al.}(2012)\citenamefont {Perra},
  \citenamefont {Gon{\c{c}}alves}, \citenamefont {Pastor-Satorras},\ and\
  \citenamefont {Vespignani}}]{Perra}%
  \BibitemOpen
  \bibfield  {author} {\bibinfo {author} {\bibfnamefont {N.}~\bibnamefont
  {Perra}}, \bibinfo {author} {\bibfnamefont {B.}~\bibnamefont
  {Gon{\c{c}}alves}}, \bibinfo {author} {\bibfnamefont {R.}~\bibnamefont
  {Pastor-Satorras}}, \ and\ \bibinfo {author} {\bibfnamefont {A.}~\bibnamefont
  {Vespignani}},\ }\href@noop {} {\bibfield  {journal} {\bibinfo  {journal}
  {Scientific Reports}\ }\textbf {\bibinfo {volume} {2}},\ \bibinfo {pages}
  {469} (\bibinfo {year} {2012})}\BibitemShut {NoStop}%
\bibitem [{\citenamefont {Valdano}\ \emph {et~al.}(2015)\citenamefont
  {Valdano}, \citenamefont {Ferreri}, \citenamefont {Poletto},\ and\
  \citenamefont {Colizza}}]{Valdano}%
  \BibitemOpen
  \bibfield  {author} {\bibinfo {author} {\bibfnamefont {E.}~\bibnamefont
  {Valdano}}, \bibinfo {author} {\bibfnamefont {L.}~\bibnamefont {Ferreri}},
  \bibinfo {author} {\bibfnamefont {C.}~\bibnamefont {Poletto}}, \ and\
  \bibinfo {author} {\bibfnamefont {V.}~\bibnamefont {Colizza}},\ }\href@noop
  {} {\bibfield  {journal} {\bibinfo  {journal} {Physical Review X}\ }\textbf
  {\bibinfo {volume} {5}},\ \bibinfo {pages} {021005} (\bibinfo {year}
  {2015})}\BibitemShut {NoStop}%
\bibitem [{\citenamefont {Cellai}\ and\ \citenamefont
  {Bianconi}(2016)}]{cellai2016}%
  \BibitemOpen
  \bibfield  {author} {\bibinfo {author} {\bibfnamefont {D.}~\bibnamefont
  {Cellai}}\ and\ \bibinfo {author} {\bibfnamefont {G.}~\bibnamefont
  {Bianconi}},\ }\href@noop {} {\bibfield  {journal} {\bibinfo  {journal}
  {Physical Review E}\ }\textbf {\bibinfo {volume} {93}},\ \bibinfo {pages}
  {032302} (\bibinfo {year} {2016})}\BibitemShut {NoStop}%
\bibitem [{sof()}]{software}%
  \BibitemOpen
  \href@noop {} {\enquote {\bibinfo {title}
  {\url{https://github.com/ikryven/InterdependentPercolation}},}\ }\BibitemShut
  {NoStop}%
\end{thebibliography}%

\end{document}